\documentclass[cits]{PoS}

\pdfoutput=1

\def\tr{\mbox{tr }}

\title{Exploring the QCD vacuum --\\
(some) recent developments in confinement and topology}

\ShortTitle{QCD vacuum}

\author{\speaker{Falk Bruckmann}\\
        Institut f\"ur Theoretische Physik, Universit\"at Regensburg, D-93040 Regensburg, Germany\\
        E-mail: \email{falk.bruckmann@physik.uni-r.de}}

\abstract{I start by giving a brief overview over new developments in the area of confinement and topology. As an example for the interrelations between topological objects, instantons at finite temperature are discussed. Then I focus on new insights into the structure of the QCD vacuum obtained through filtering methods, in particular 
those based on decompositions w.r.t.\ fermionic eigenmodes.}

\FullConference{The XXV International Symposium on Lattice Field Theory\\
		 July 30 - August 4, 2007\\
		 Regensburg, Germany}

\begin{document}

\section{Confinement: areas of activity not covered}

Confinement, one of the main phenomena in QCD, 
is usually associated with a gluonic string between quarks and antiquarks giving rise to a linear potential.
This QCD string remains under investigation on the lattice, 
in particular its breaking by dynamical quarks has been seen in lattice simulations \cite{bali:05}.

\begin{figure}[h]
\begin{center}
\includegraphics[width=0.6\linewidth]{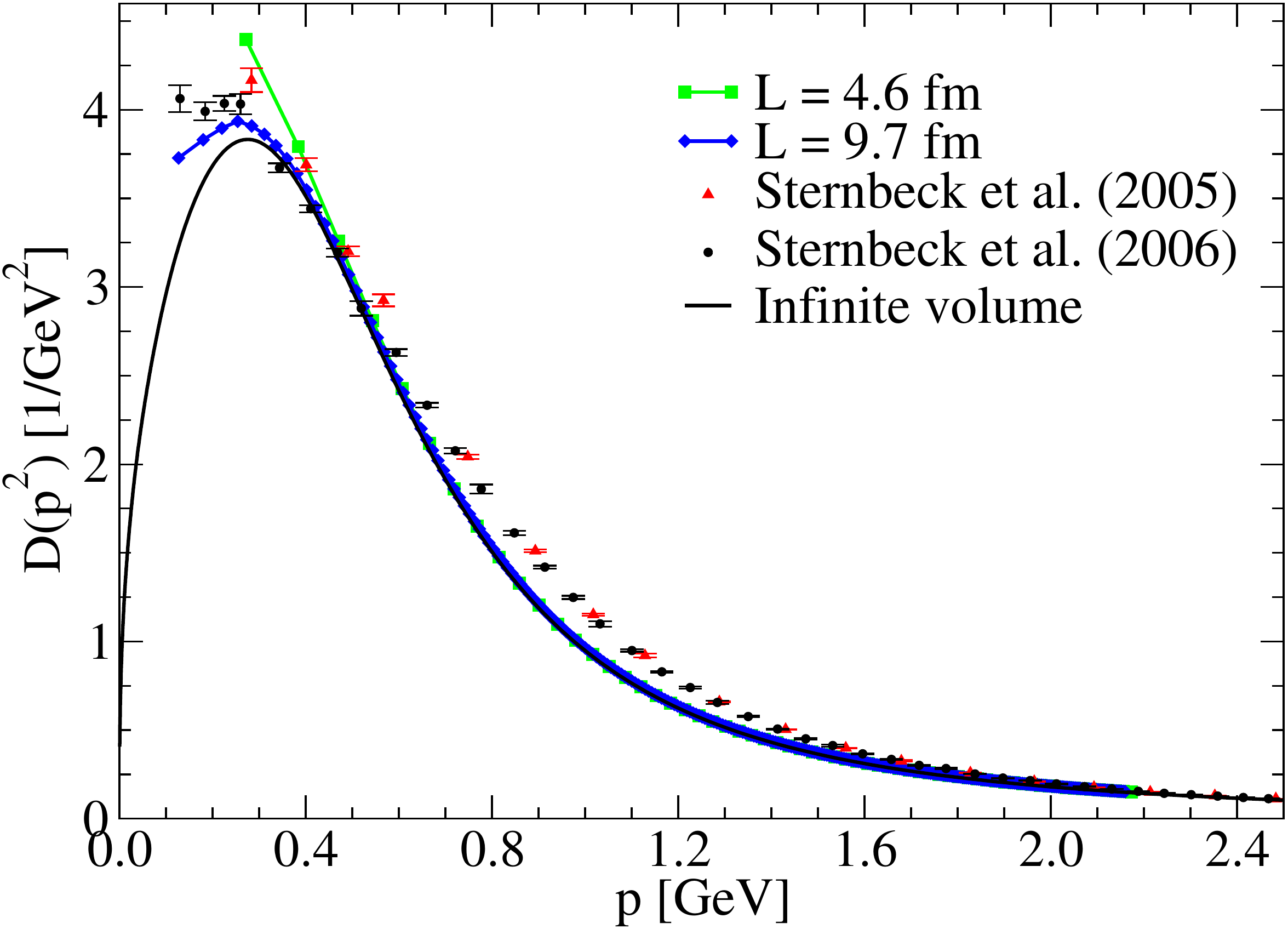}
\caption{Comparison of Dyson-Schwinger (curves) and lattice results (data points)
on the gluon propagator in Landau gauge, from ref.~\protect\cite{fischer:07a}.}
\end{center}
\label{fig_prop}
\end{figure}

However, there are other criteria for confinement that 
dominated this years confinement session. 
The Kugo-Ojima and Gribov-Zwanziger confinement criterion concern the infrared properties of propagators in Landau gauge. 
According to these criteria, the gluon propagator behaves as 
\begin{equation}
D(p^2)\stackrel{p\to 0}{\longrightarrow} (p^2)^{2\kappa-1},\qquad \kappa>0\,.
\end{equation} 
Dyson-Schwinger equations in the continuum predict $\kappa  = 0.595$, but
with finite volume effects up to $L\gg 2\pi/\Lambda_{QCD}\simeq 5$ fm.
Fig.\ \ref{fig_prop} shows how the finite volume curves bend towards zero as $p\rightarrow 0$,
together with some first lattice computations of 
this observable.
This figure is taken from ref.\ \cite{fischer:07a}, 
where many more references and details of the approach can be found.
At this conference, various groups presented lattice computations on huge lattices, 
but so far could not grasp beyond the plateau in the gluon propagator.
This mismatch needs to be clarified.

A partially similar method utilizes the Schr\"odinger picture in Coulomb gauge.
In both cases gauge fixing is plagued by the Gribov problem. 
The latter is rooted in the non-trivial configuration space of gauge theories.
It is not always a problem, 
but has also been used to get information about the equation of state \cite{zwanziger:04}. 
A completely different approach to confinement 
is based on renormalization group decimation \cite{tomboulis:07}.

\pagebreak

\section{Topology}

The  question which nonperturbative degrees of freedom are relevant for the QCD vacuum 
brings me to the topic of topology. 
Over the years there has been established a `Standard model' of topological objects. 
Table \ref{tab_top_objects} summarizes properties of center vortices, magnetic monopoles and instantons, 
namely their dimensionality, 
to which subgroup of the gauge group they are related 
and by which technqiue they can be obtained, e.g. on the lattice. 

\begin{table}[h]
\begin{center}
\begin{tabular}{r@{\extracolsep{-0.0cm}}c|@{\extracolsep{0.2cm}}c@{\extracolsep{1cm}}c@{\extracolsep{1cm}}c}
&& center vortices             &    magnetic monopoles        &    instantons               \\
\hline
dimensionality && 2 dim.\ sheets               &    1 dim.\ worldlines                &    pointlike                \\
subgroup && center                      &    max.\ Abelian subgroup    &    full gauge group         \\
technique && center projection           &    Abelian projection     &    semiclassics 
\end{tabular}
\caption{Properties of topological degrees of freedom proposed in the QCD vacuum.}
\label{tab_top_objects}
\end{center}
\end{table}

The effect of chiral symmetry breaking is rather robust 
in that all these topological excitations have been shown to induce it.
Confinement is generated by vortices and monopoles, 
but not by instantons;
for a detailed review and a discussion of the interesting case of a centerless gauge group
see the plenary talks of M.\ Engelhardt \cite{engelhardt:04a} 
and M. Pepe \cite{pepe:05} at previous Lattice conferences.

I would like to emphasize some new insights about interrelations of these objects. 
Concerning vortices and monopoles, it has been observed 
that removing one type of object destroys the confinement mechanism of the other \cite{boyko:06}.
Monopoles and instantons, on the other hand,  are related at finite temperature, 
which will be discussed in more detail in the next section.

\section{Calorons}

Instantons are configurations with minimal action in a sector of non-trivial topological charge. 
They are classical solutions, fulfilling in addition first oder (self-duality) equations.
In most cases, instantons are localised lumps of action/topological density
with locations, sizes and color orientations as moduli.

\textit{Calorons} are instantons at finite temperature. 
In the continuum they are defined over the manifold $S^1\times\mathbb{R}^{d-1}$ 
whereas on the lattice on takes one extension much smaller than the others, $L_0\ll L_i$;
the circumference of the compact direction is $\beta=1/k_BT$.

Calorons can be constructed naturally by putting infinitely many instanton copies along $x_0$. 
This leads to overlap effects depending on the size and the relative color orientation of these copies.

\begin{figure}[t]
\includegraphics[width=0.65\linewidth]{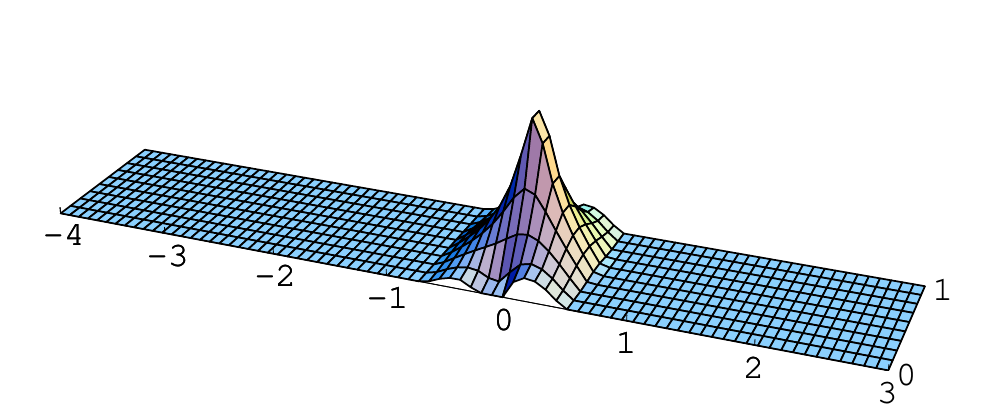}

\vspace{-5.4cm}\hspace*{5.5cm}
\includegraphics[width=0.65\linewidth]{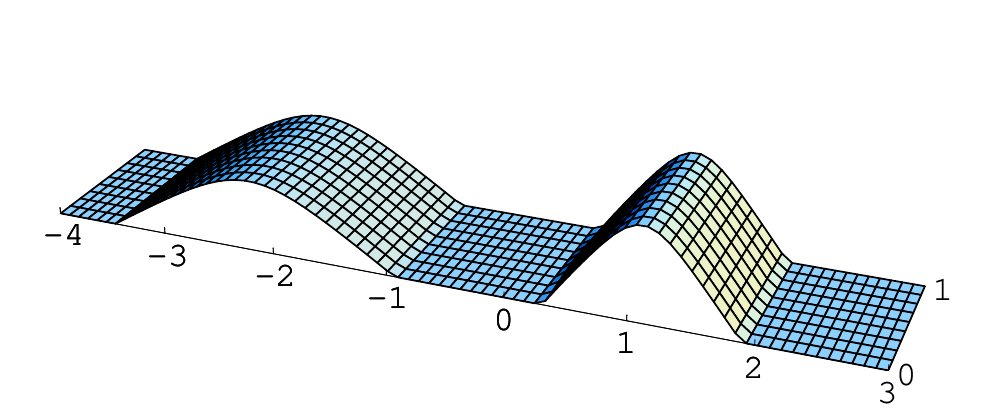}

\vspace{1.3cm}
\caption{The topological charge of calorons in the $O(3)$ model, 
eq.s (\protect\ref{eqn_u}) and (\protect\ref{eqn_o3_q}),
with holonomy parameter $\omega=1/3$ 
and small (left) and large (right) size para\-meter $\lambda$ 
($0.03$ and $100$ in units of $\beta$, the extension of the compact direction),
cf.\ ref. \protect\cite{bruckmann:07b}.
.}
\label{fig_profiles_o3}
\end{figure}

Let me illustrate this with a rather simple example, the two-dimensional $O(3)$ model.
Identifying $\mathbb{R}^2$ with the complex plane, it can be shown 
that any meromorphic function is a solution of the first order equations 
and hence 
of the equations of motion of this system. 
The pole $z_0$ in the simplest meromorphic functions,
\begin{equation}
u(z)=(z-z_0)/\lambda\,,\qquad u(z)=\lambda/(z-z_0)\,,\qquad \lambda,z_0\in\mathbb{C}\,,
\end{equation} has the meaning of the center of the instanton, 
whereas $|\lambda|$ is its size.

The conjecture that in this model instantons may contain a substructure,
known under the name of `instanton quarks',
arose from an alternative rational parametrisation,
\begin{equation}
u(z)=\frac{z-\hat{z}}{z-\check{z}}\,,\qquad \hat{z},\check{z}\in\mathbb{C}. 
\end{equation}
However, also this $\mathbb{R}^2$ instanton consists of one lump, centered at $(\hat{z}+\check{z})/2$
with size $|\hat{z}-\check{z}|/2$.

Recently I have analyzed the finite temperature case \cite{bruckmann:07b}. The function 
\begin{equation}
u(z)=\frac{\lambda\cdot e^{\omega \frac{2\pi z}{\beta}}}{e^{\frac{2\pi z}{\beta}}-1}\,,\qquad \omega\in [0,1]\,,
\label{eqn_u}
\end{equation}
has poles at $z=k\cdot i\beta$ with residues $\lambda\cdot e^{2\pi i\omega k}$.
Hence it realizes a caloron coming from instanton copies at $x_1\equiv \mbox{Re }z=0$ of size $|\lambda|$ 
and with orientations constantly rotated by $e^{2\pi i\omega}$. 

The topological charge density in this model has a rather 
simple form in terms of $u(z)$ and its derivative:
\begin{equation}
q=\frac{1}{\pi}\frac{1}{(1+|u|^2)^2}\left|\frac{\partial u}{\partial z}\right|^2
\label{eqn_o3_q}\,.
\end{equation}
Fig.\ \ref{fig_profiles_o3} shows profiles of $q$
for two caloron solutions with non-trivial parameter $\omega$ in (\ref{eqn_u}). 
While the small caloron resembles a pointlike instanton, 
the large caloron consists of {\em static constituents with fractional charges 
$\omega$ and $1-\omega$}, adding up to one unit.
 The size parameter of large calorons transmutes into the distance of its constituents,
which themselves have a size proportional to $\beta$.
The role of these constituents in the dynamics of the $O(3)$ model at finite temperature has not been explored yet.\\


In 3+1 dimensional (pure) gauge theories one needs the ADHM formalism 
to construct the most general caloron solution. 
Again, large calorons dissociate into static constituents, 
namely {\em magnetic monopoles} \cite{kraan:98a,lee:98b}.
This and other features have made the caloron attractive in recent years.

\begin{figure}[t]
\begin{center}
\includegraphics[width=0.55\linewidth]{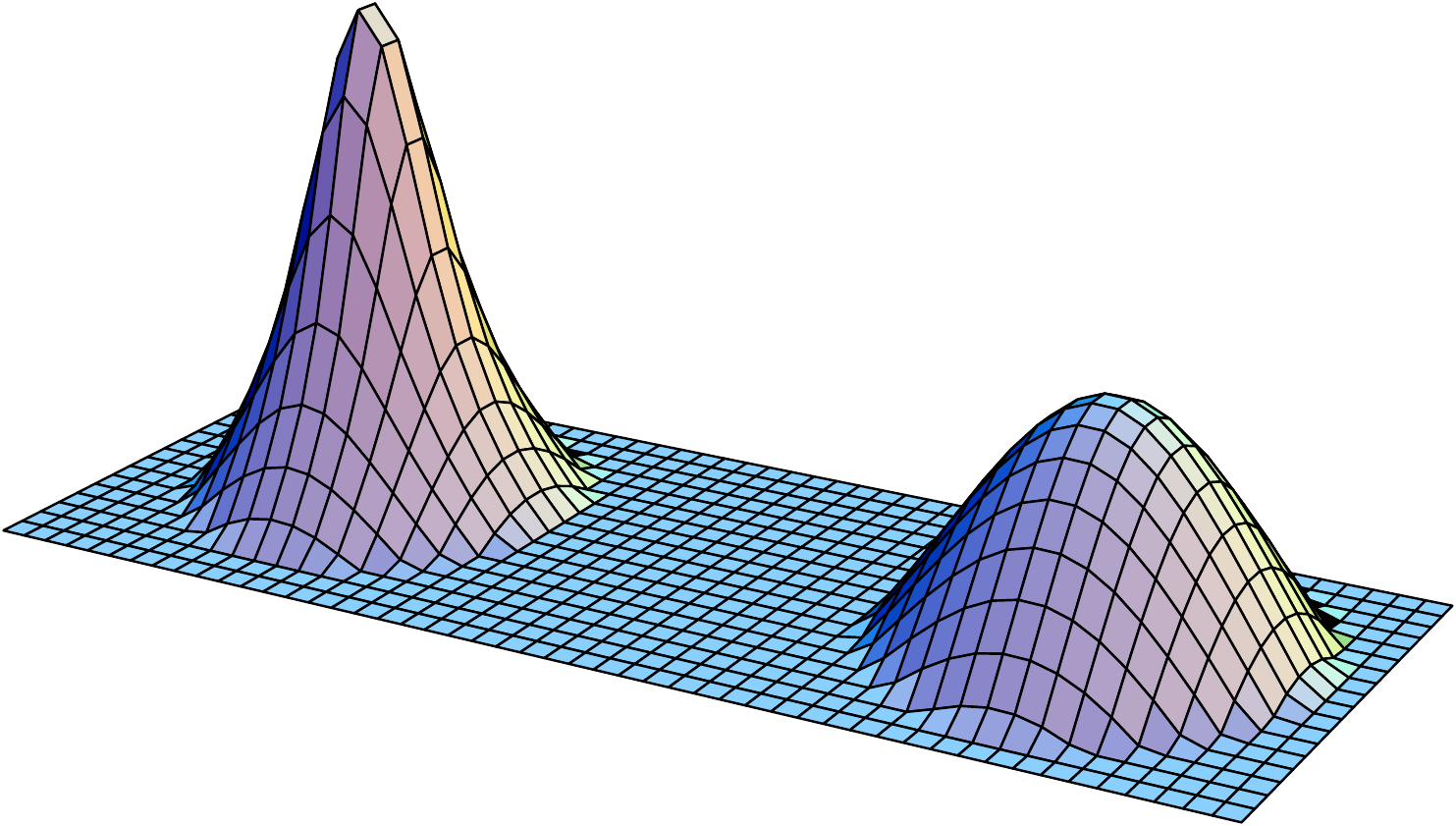}
\caption{Space-space plot of an $SU(2)$ caloron with $\omega=1/8$ and size 1.6\,$\beta$, from \protect\cite{kraan:98a}.}
\label{fig_profile_caloron}
\end{center}
\end{figure}

Fig.\ \ref{fig_profile_caloron} shows such a `dissociated' caloron.
For gauge group $SU(2)$, the monopole masses are $2\omega$ and $1-2\omega$.
In the limiting case of $\omega=0$ one of the monopoles is massless 
and one arrives at the old Harrington-Shephard caloron \cite{harrington:78}.

For gauge group $SU(N_c)$ there are $N_c$ constituent monopoles, 
just like quarks in a baryon. 
Due to self-duality, these objects are electrically charged as well
and therefore can be called dyons, too. 
Some higher charge calorons are known \cite{bruckmann:02b,bruckmann:03a,bruckmann:04a}
with the number of consituents of each type being proportional to the topological charge.
The constituent monopoles also carry the fermionic zero modes
(depending on the boundary conditions),
an interesting story on its own.

By construction the gauge fields are periodic up to a gauge transformation.  
They can be made periodic by a time-dependent gauge transformation, 
which in turn gives rise to an asymptotic field $A_0(|\vec{x}|\to\infty)$.
Accordingly, the asymptotic Polyakov loop, named holonomy, becomes non-trivial
\begin{equation}
\mathcal{P}_\infty\equiv \lim_{|\vec{x}|\to\infty}\mathcal{P}\exp\int_0^\beta A_0\, d x_0
= e^{\,2\pi i\omega\sigma_3}\,.
\end{equation}
It serves as a Higgs field (in the gauge group) 
or as a non-trivial `background' for this novel type of instanton solution.
$\omega$ is the vacuum expectation value governing the masses of the monopoles, 
while their charges appear in the long-range gauge field along the Higgs direction $\sigma_3$
(in contrast to these `photon' fields, the `$W$-boson' fields proportional to $\sigma_{1,2}$ decay exponentially).

Apparently,
the instanton liquid model at finite temperature needs to be modified.
The hope is to make contact to the dual superconductor picture based on magnetic monopoles and
to the deconfinement order parameter.
In the confined phase the latter is $\langle\tr\,\mathcal{P}\rangle=0$, 
which would be best mimicked by $\tr\:\mathcal{P}_\infty=0$, the case of 
maximally non-trivial holonomy with identical monopoles.

Two recent findings are very encouraging: 
The first one concerns the effective potential at finite temperature
for the gauge field in the compact direction. 
It is well-known that at one loop the effective potential favours trivial holonomy $\omega=0$ (or $\omega=1/2$).
This argument, however, is overruled by a nonperturbative contribution from a caloron gas
\cite{diakonov:04a}:
lowering the temperature, $\omega=0$ becomes unstable and non-trivial holonomy is preferred,
which can be interpreted as the onset of confinement, visible in the caloron properties.

The second investigation is a semi-analytical study of calorons at $T\simeq T_c$ \cite{gerhold:06}. 
Calorons of fixed holonomy have been superposed and discretized on a lattice. 
As the main effect, a linearly rising interquark potential has been found just for non-trivial holonomy.

\section{Filtering methods}

The identification of topological objects in the QCD vacuum, 
as represented by lattice configurations,  
is plaqued by the fact that the latter are dominated by  UV (i.e.\ $O(a)$) fluctuations. 
Therefore I will now discuss methods to filter these lattice configurations.

The most well-known filtering methods are cooling and smearing\footnote{
I will always refer to APE smearing, equivalent to RG cycling \cite{degrand:98b}. 
Its parameters in eq. (\ref{eqn_smearing}) are $\alpha=0.55$ and $\gamma=0.075$,  while for cooling $\alpha=0$.}, 
iterative modifications of the links that average them with the surrounding staples $U_\mu^{(\nu)}(x)$,
\begin{equation}
U_\mu(x)\to P\big[\alpha U_\mu(x)+\gamma\sum_{\nu\neq \mu} U_\mu^{(\nu)}(x)\big]\,,\qquad \alpha,\gamma\in[0,1]\,.
\label{eqn_smearing}
\end{equation}
In the case of $SU(2)$ the projection $P$\, back onto the gauge group is just a scalar multiplication.

This method removes the UV `noise' and drives the configuration towards classical solutions. 
For example, calorons on the lattice have been obtained this way \cite{ilgenfritz:02a,ilgenfritz:05}.
As a filtering method, however, smearing is biased, it prefers instantons (of certain size). 
Moreover, it is not clear how long smearing should be applied.\\

Recently developed filtering methods are based on the eigenvalues and eigenvectors of lattice Dirac operators.
One of the key ideas is that (isolated) vortices, monopoles and instantons possess a topological zero mode.
Hence, low-lying eigenmodes are expected to be localized to topological objects.

This localization has been used to probe the dimensionality of the underlying topological excitations 
by the scaling of the inverse participation ratio with the lattice spacing \cite{aubin:04,gubarev:05a}.
Some evidence for brane-like objects was reported, 
but altogether the approach seems inconclusive \cite{deforcrand:06}.

By virtue of chiral fermions the topological charge density can be defined on the lattice 
alternatively to gluonic definitions. 
With a Ginsparg-Wilson type Dirac operator $D$ it reads \cite{niedermayer:98}:
\begin{equation}
q(x)\equiv\tr\, \gamma_5(\frac{1}{2}\,D_{x,x}-1)
=\sum_{n=1}^{Vol\cdot 4 N_c} (\frac{\lambda_n}{2}-1)\psi^\dagger_n(x)\gamma_5\psi_n(x)\,.
\label{eqn_q_ferm}
\end{equation}
$q(x)$ sums up to an integer, determined purely by the zero modes. 
Moreover, this definition approaches the conventional $\tr\,F_{\mu\nu}\tilde{F}_{\mu\nu}/16\pi^2$ for smooth configurations.

In eq. (\ref{eqn_q_ferm}) I have included the spectral representation of the topological charge density.
Now the idea of filtering is to {\em truncate the corresponding spectral sum} at a rather small number $N \ll Vol$ of modes.
$\lambda_N$ can then be viewed as the resolution of the filter.

\begin{table}[t]
\begin{center}
\begin{tabular}{c@{\extracolsep{\fill}}|cccccc}
 $q_{\mbox{{\small cut}}}/q_{\mbox{{\small max}}}$  &0.00&0.10&0.20&0.30&0.40&0.50\\
\hline
 all-scale &2.7(1)&1.5(1)&1.1(1)&0.7(1)&0.4(1)&0.1(1)\\
$\lambda_{\mbox{cut}}\le$ 200 MeV $\:$&3.1(1)&1.8(1)&1.5(1)&1.0(1)&0.7(1)&0.2(1)\\
\end{tabular}
\end{center}
\caption{Dimensions of topological structures found by truncating the fermionic definition 
of the topological charge density, eq. (\protect\ref{eqn_q_ferm}), at some eigenvalue
and cutting off the absolute value of $q$ at some level, from ref.\ \protect\cite{ilgenfritz:07a}.
Note the appearance of (roughly) three-dimensional objects at low cut-off.}
\label{tab_dims}
\end{table}

\begin{figure*}[b]

\begin{minipage}{0.24\linewidth}
\hspace{-0.8cm}
\includegraphics[width=1.15\linewidth]{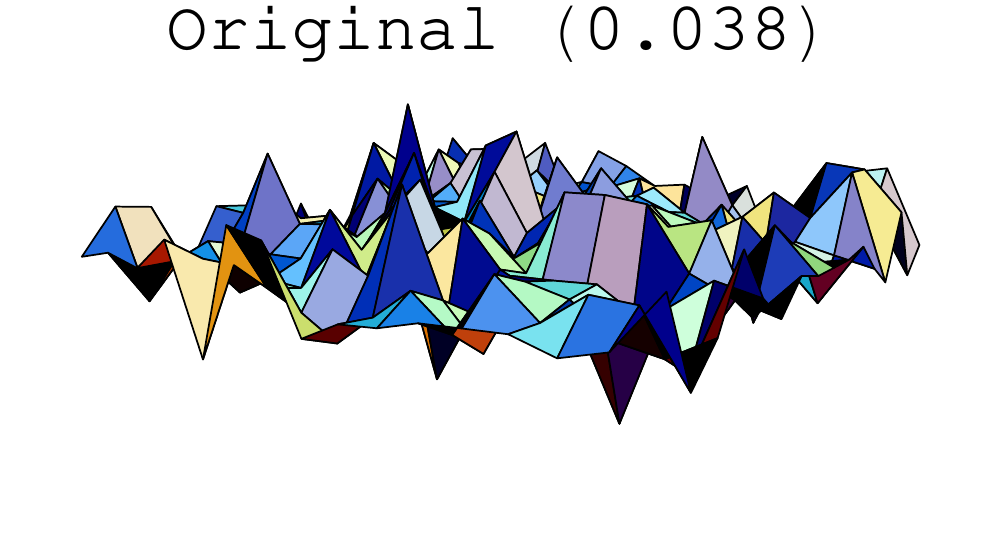}
\end{minipage}
\begin{minipage}{0.72\linewidth}
\hspace{-0.7cm}
\includegraphics[width=0.385\linewidth]{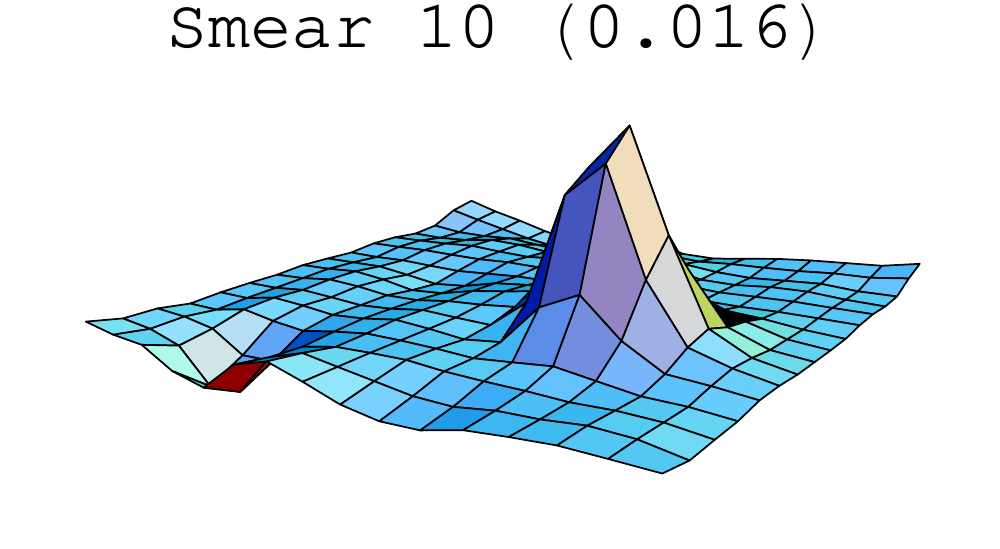}
\hspace{-0.7cm}
\includegraphics[width=0.385\linewidth]{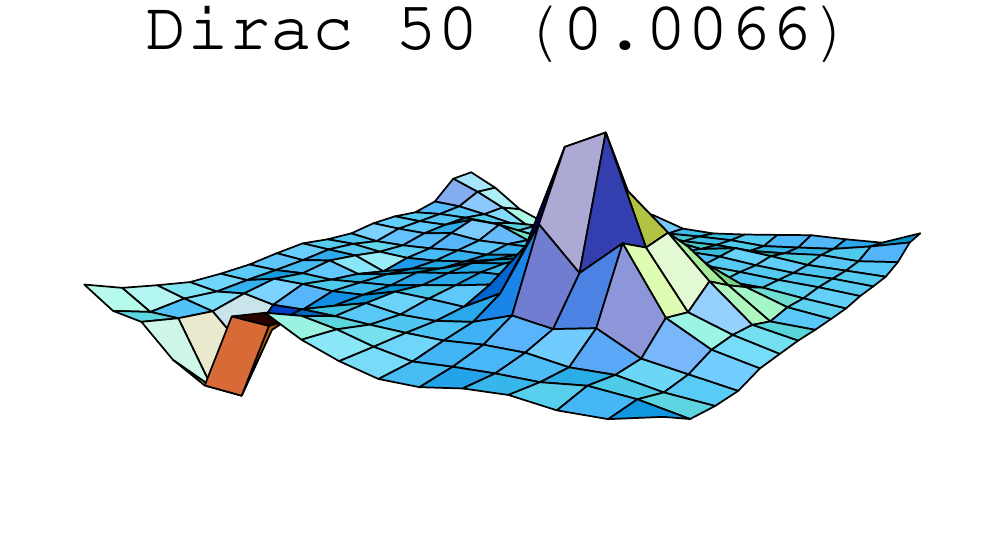}
\hspace{-0.7cm}
\includegraphics[width=0.385\linewidth]{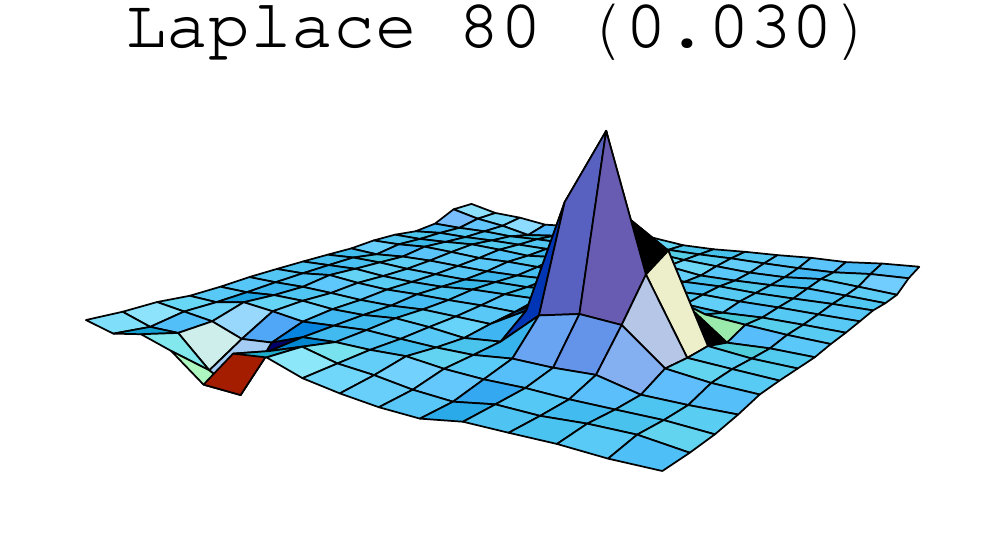}
\\

\hspace{-0.7cm}
\includegraphics[width=0.385\linewidth]{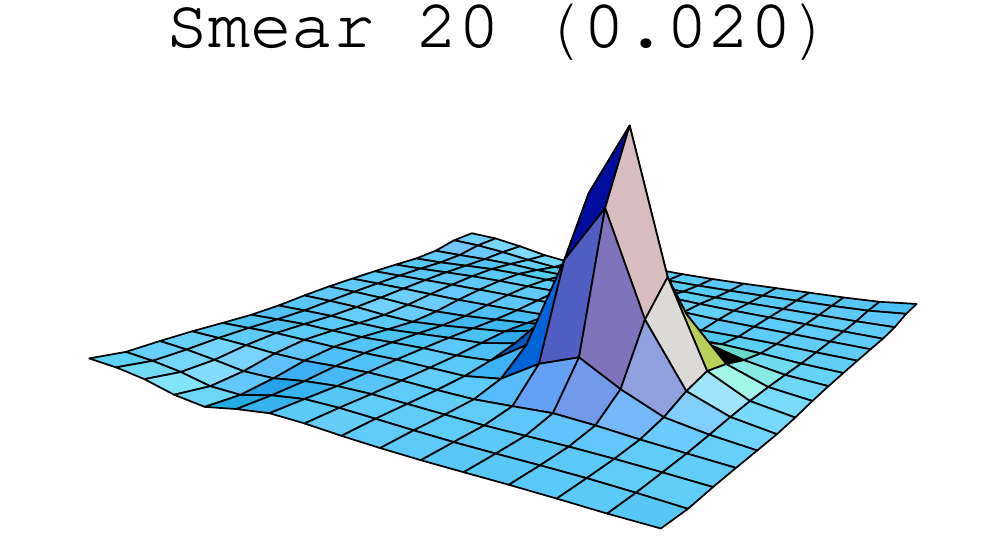}
\hspace{-0.7cm}
\includegraphics[width=0.385\linewidth]{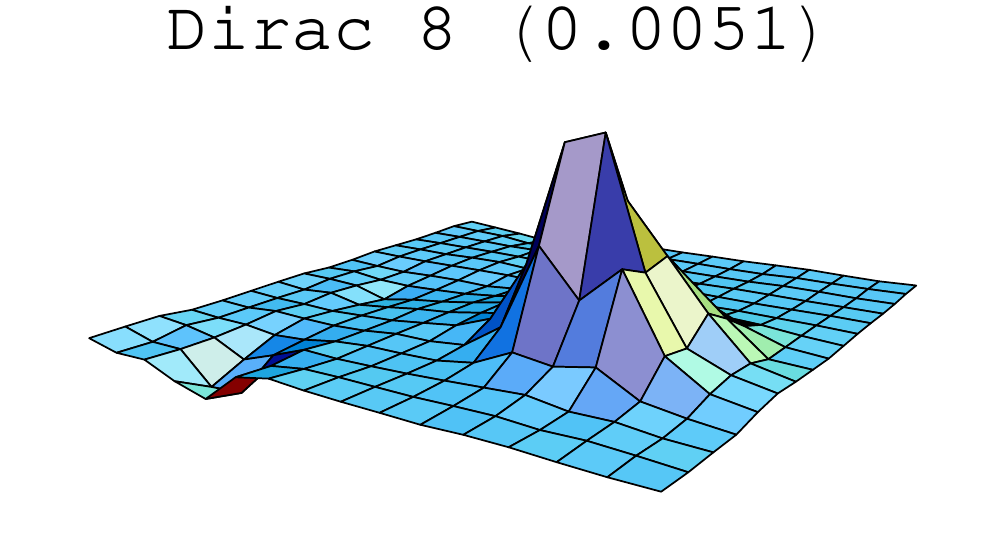}
\hspace{-0.7cm}
\includegraphics[width=0.385\linewidth]{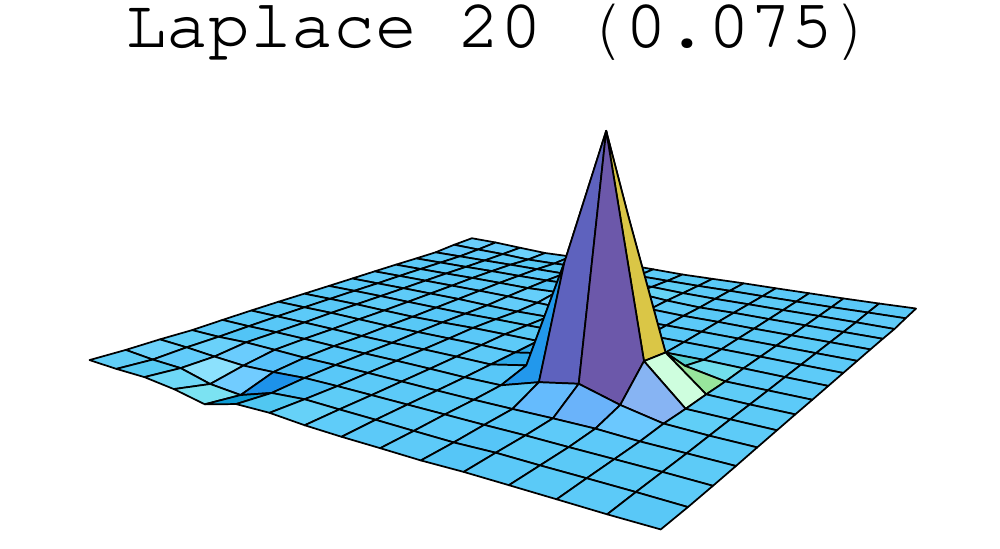}
\end{minipage}
\caption{The topological charge density of a thermalized (quenched $SU(2)$ zero temperature)
configuration in a fixed lattice plane, from ref.~\protect\cite{bruckmann:06a}.
The left panel shows the original configuration with the typical UV fluctuations.
The other panels show the filtered topological charge density after smearing, 
Dirac and Laplace filtering 
(according to eq.s (\protect\ref{eqn_smearing}), (\protect\ref{eqn_q_ferm}) and (\protect\ref{eqn_laplace}))
at different levels of filtering (top = mild, bottom = strong filtering, cf.\ Fig.~\protect\ref{fig_matching}).} 
\label{fig_showcase_plane}
\end{figure*}

\begin{figure}[t]
\begin{center}
\includegraphics[width=0.32\linewidth]{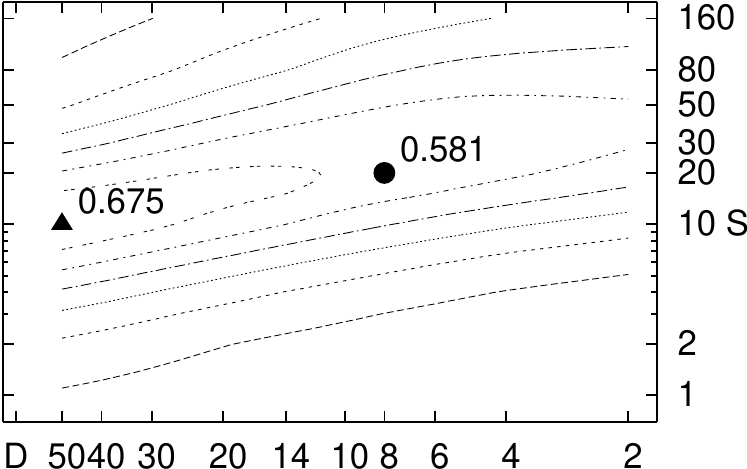}\hfill
\includegraphics[width=0.32\linewidth]{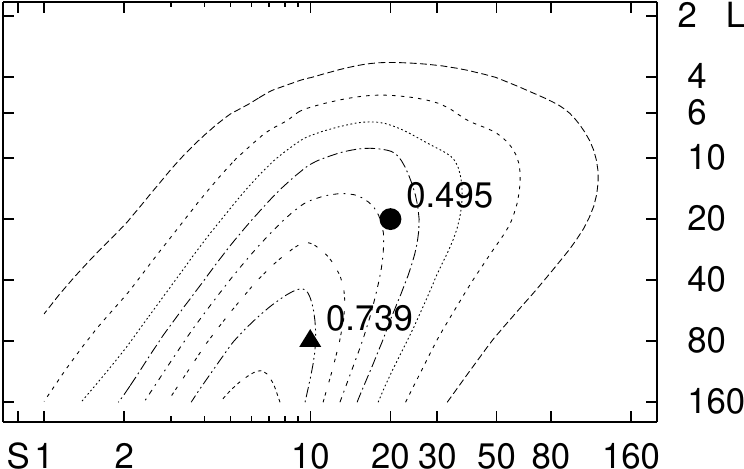}\hfill 
\includegraphics[width=0.32\linewidth]{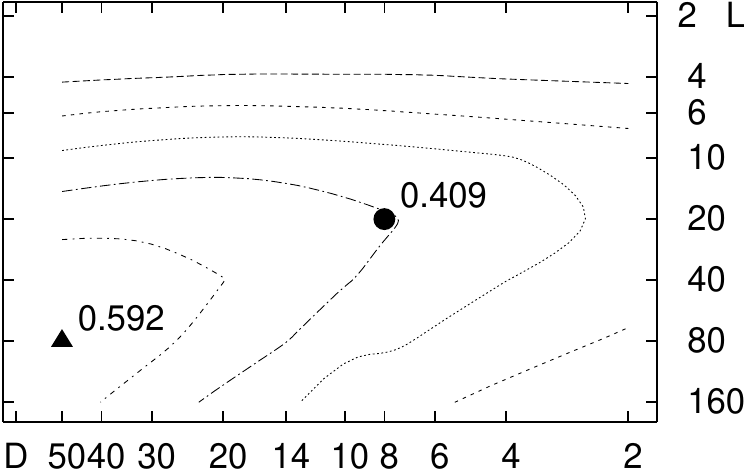}
\caption{Matching of parameters of different filtering methods by maximizing $\Xi_{AB}$, 
eqn. (\protect\ref{eqn_correlators}).
S stands for smearing, D for the fermionic topological charge and L for the Laplace filter,
respectively, each with the corresponding number of sweeps, fermionic or Laplacian modes.
The triangles denote mild filtering, the dots stronger filtering.}
\label{fig_matching}
\end{center}
\end{figure}

In this way, first evidence for extended three-dimensional structures has been found by Horvath et al.\ \cite{horvath:03a}.
A refined analysis revealed all lower dimensions depending on the cut-off (and the resolution), 
see table \ref{tab_dims} \cite{ilgenfritz:07a}.
A peculiarity emerges at very low cut-off: 
space-time gets filled by two connected components of topological charge,
one positively and one negatively charged, being entangled such 
that the opposite topological charge is nowhere far away.\\

\vspace{1cm}

With the different filtering methods at hand, the obvious question is whether
they\footnote{including the Laplace filter to be described below} 
reveal the same topological structures. 
As the visualization in fig. \ref{fig_showcase_plane} suggests, 
all methods find the same `hot spots' of topological charge \cite{bruckmann:06a}. 
This excellent agreement is highly non-trivial given the different nature of the filtering methods.

For the matching of the different filter parameters (and a more quantitative comparison)
we have considered crosscorrelators
\begin{equation}
\chi_{AB}(r)\equiv\langle q_A(0)q_B(r)\rangle\,,\qquad
\Xi_{AB}\equiv\frac{\chi_{AB}^2(0)}{\chi_{AA}(0)\chi_{BB}(0)}\,.
\label{eqn_correlators}
\end{equation}
The quantity $\Xi_{AB}$ is 1, if both methods $A$ and $B$ give identical results. 
Fig.\ \ref{fig_matching} demonstrates how an optimization of $\Xi_{AB}$ 
serves to match the parameters of the filtering methods \cite{bruckmann:06a}.
Structures identified in this way faithfully represent infrared degrees of freedom in the QCD vacuum.

The distribution of the filtered topological lumps, in particular those which are common to all methods,
reveals an interesting power-law. 
It can be used to exclude a dilute instanton gas as a model for the structures after filtering
(for more details see S. Solbrig's talk \cite{solbrig:07a}).

\vspace{1cm}

\section{More spectral decompositions}

In the comparison of the filtering methods, the Laplace filter \cite{bruckmann:05b}
has been included, which will be discussed now.
From the definition of the gauge-covariant lattice Laplace operator 
and its spectral decomposition,
\begin{equation}
-\Delta[U]^{ab}_{xy}\equiv\sum_\mu [-U_\mu^{ab}(x)\delta_{x+\hat{\mu},y}-
U_\mu^{\dagger ab}(y)\delta_{x-\hat{\mu},y}+2\delta^{ab}_{xy}]
=\sum_n^{Vol\cdot N_c}\lambda_n \phi_n^a(x)\phi^{*b}_n(y)\,,
\end{equation}
an exact formula for the gauge links in terms of all Laplacian eigenvalues and
eigenmodes follows:
\begin{equation}
U_\mu^{ab}(x)=-\sum_n^{Vol\cdot N_c}\!\!\lambda_n\, \phi_n^a(x)\phi^{*b}_n(x+\hat{\mu})\,.
\label{eqn_laplace}
\end{equation}
For filtering purposes it is again truncated at small $N$ (and projected back onto the gauge group):
\begin{equation}
U_\mu^{(N)ab}(x)=-P\,\big[\sum_n^{N}\lambda_n\, \phi_n^a(x)\phi_n^{*b}(x+\hat{\mu})\big]\qquad N\ll Vol\,.
\end{equation}

This method gives back filtered links, hence all observables can be measured on them.
It is also numerically cheaper than the fermionic filter (\ref{eqn_q_ferm}).
The string tension is preserved after filtering \cite{bruckmann:05b}, 
hence one can speak of a {\em low mode dominance} of confinement.
And as I have shown before, it agrees to a large extent with smearing, although it is in principle nonlocal.\\

Other observables that have been given in a spectral representation of some lattice operators 
are the field strength \cite{gattringer:02c,liu:07,ilgenfritz:07a} and the action \cite{gonzalez-arroyo:05}.\\ 

\begin{figure}[b]
\begin{center}
\includegraphics[width=0.4\linewidth]{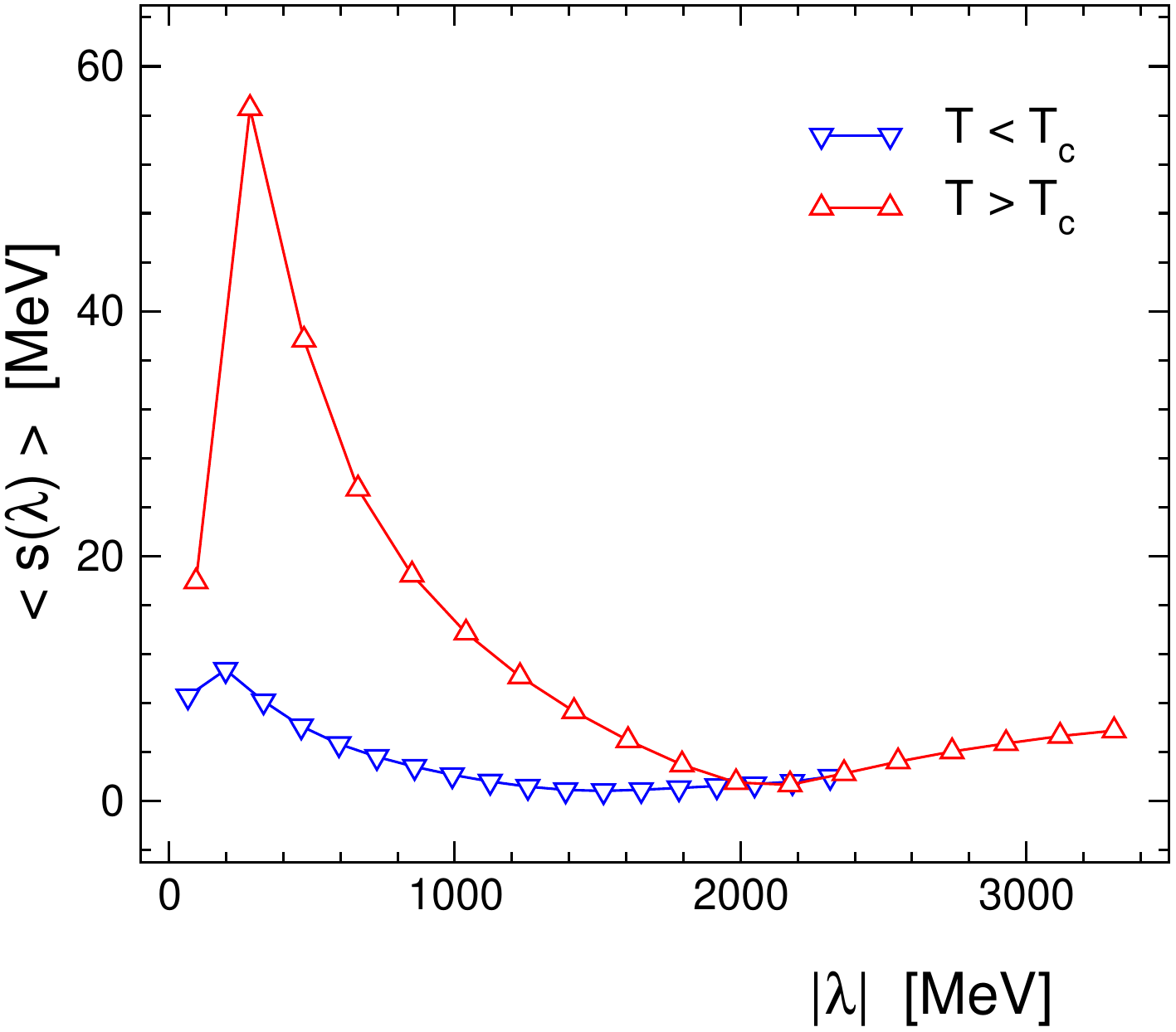}
\qquad
\includegraphics[width=0.4\linewidth]{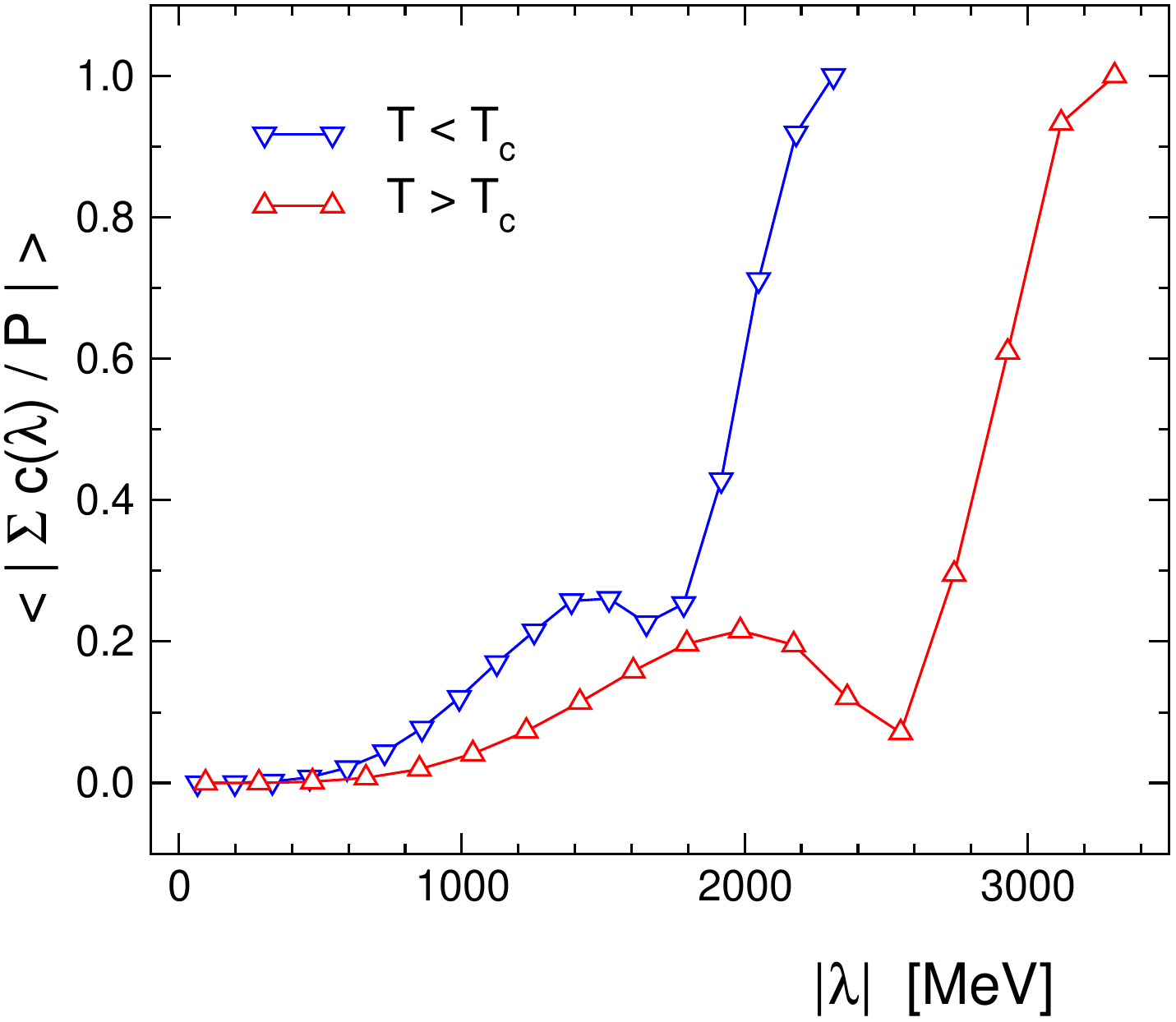}
\caption{Details of the reconstruction 
of the Polyakov loop through eigenvalues of the staggered (quenched)
$SU(3)$ Dirac operator on a $6^3\cdot 4$ lattice, 
from \protect\cite{bruckmann:06b}.
The left panel shows a quantity measuring the change of the eigenvalue 
with the 3 boundary conditions in eq.\ (\protect\ref{eqn_polloop}).
The right panel displays how the absolute value of the reconstructed Polyakov loop 
(i.e.\ the r.h.s.\ of that equation) evolves to the full original Polyakov loop, set to 1.}
\label{fig_polloop}
\end{center}
\end{figure}

Of particular interest is a representation of the Polyakov loop, 
as the latter encodes the (de)con\-fi\-ning properties of QCD.
This method starts with the fact
that products of lattice operators $D$ generate products of links.
Taken at same initial and final point, these products induce closed loops.
In the spectral representation of these loops the eigenvalue 
is simply raised to the according power in the product.

The other main ingredient is that different phase boundary conditions in the temporal direction 
can be used to distinguish Polyakov loops from `trivially closed' loops (like the plaquette) \cite{gattringer:06b}.
Summing over all space points, the following formula is exact:
\begin{equation}
\sum_x\tr\mathcal{P}(x)=
\mbox{const.} \,\sum_n^{Vol\cdot N_c}\big[(\lambda_n)^{N_t}+z^*\,(\lambda_{z,n})^{N_t}+z\,(\lambda_{z^*,n})^{N_t}\big]\,.
\label{eqn_polloop}
\end{equation}
Here, $\lambda_z$ are the eigenvalues of the staggered Dirac operator $D$ -- other one-link operators can be used as well -- with boundary conditions chosen to be $z=\exp(2\pi i/3)$, 
the gauge group is $SU(3)$. 

This relation has the potential to describe (de)confinement in terms of Dirac spectra, 
which is very attractive since chiral symmetry breaking is related to the eigenvalue density at $\lambda=0$ via the Banks-Casher relation.

For filtering purposes, the sum on the r.h.s.\ of eq.\ (\ref{eqn_polloop}) can be truncated in the IR.
Generically, this will reconstruct only a fraction of the Polyakov loop, 
depending not only on the absolute value (to the $N_t$th power) 
of the eigenvalues, 
but also on their dependence on the boundary condition $z$.

As fig.\ \ref{fig_polloop} shows, the lowest modes are responding most to changing $z$, 
but contribute least to $\mathcal{P}$ \cite{bruckmann:06b}.
In other words, the (thin) Polyakov loop is obtained mainly from the UV modes. 
This remains true for other values of $N_t$ and other spatial volumes, 
for details see the talk of C.\ Hagen and the poster of E.\ Bilgici \cite{bilgici:07}
(and also for dynamical fermions \cite{soeldner:07}). 

\begin{figure}[t]
\begin{center}
\includegraphics[width=0.7\linewidth]{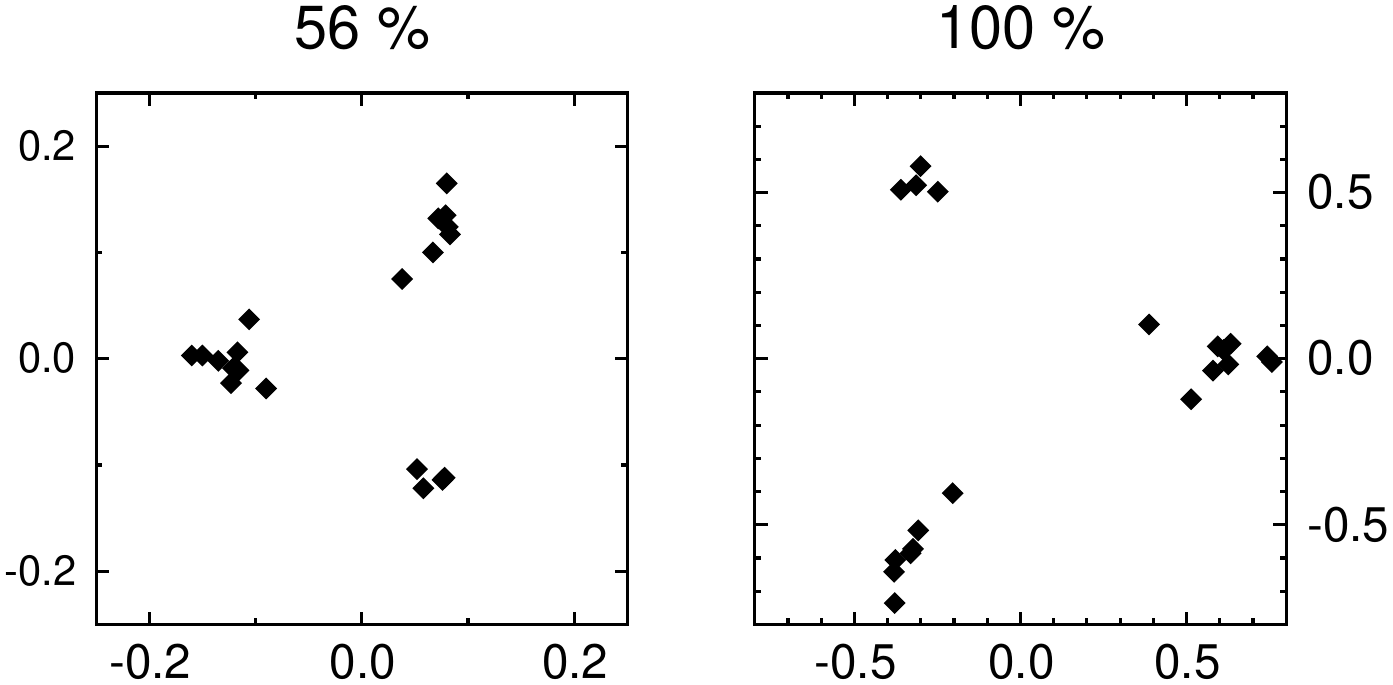}
\caption{The average Polyakov loops of 20 
independent configurations above the critical temperature in the complex plane, 
from ref.~\protect\cite{bruckmann:06b}.
Right the original Polyakov loops with the typical pattern close to the center of $SU(3)$.
Left the reconstructed Polyakov loops with 56\% 
of the eigenvalues included in the sum in eq.\ (\protect\ref{eqn_polloop}).
The latter are roughly the former reflected at the origin (and multiplied by some constant).}
\label{fig_complex}
\end{center}
\end{figure}

Why the modes at the lattice cut-off are so important is not fully understood.
Partially the issue is due to the fact that the IR modes 
have a low density (or even a gap) and a small absolute value. 
But even their small contribution comes with an overall minus sign, 
as shown in fig.\ \ref{fig_complex}.
This is caused by the twist in the boundary condition yielding larger eigenvalues 
at the lower end of the spectrum 
(for $N_t=6$ the reconstruction of the Polyakov loop starts out with the correct phase, 
but develops into the opposite phase in an intermediate region of the spectrum) \cite{bilgici:07}.

Recently, this approach has been extended to other functions of $D$ \cite{synatschke:07a}
and to dressed Polyakov loops (and the Laplacian taken for $D$) \cite{bilgici:07}.
One of the goals is to get insight into the continuum limit of this interesting connection 
between (de)confinement and chiral quantities. 

\section{Summary}

I have presented several recent investigations about the physical mechanisms in the QCD vacuum 
and the role of topological excitations therein.
So far we "tend to bear in mind that some underlying `classical' fields 
(be it `fat' monopoles, vortices or instantons) drive the phenomena.
But attempts to identify them in non-\-perturbative ensembles have sytematically led to problems. 
There might be some truth in it, but the key ingredient seems to be still missing" \cite{garciaperez:00b}.
The new results, in my opinion, exhibit the following tendencies:

The first one is towards a unification of topological objects 
in a `democratic vacuum' (containing all of them).
Calorons have been discussed as an example relating instantons and monopoles.
Three-dimensional sheets seem relevant, too.

Secondly, we have now at our disposal independent filtering methods 
to access the infrared degrees of freedom in thermalized lattice configurations.
As I have shown, they agree on the topological charge structures and can be used to distinguish 
physical excitations 
from artefacts of the methods.
Actually, most of the studies of the QCD vacuum have been performed on quenched backgrounds,
some recent ones address more realistic situations with dynamical quarks \cite{weinberg:07}.

The third tendency is that topological structures depend on cut-off and resolution.
Together with the first point, this new paradigm is a big challenge for modelling 
(also in view of the particular form of the topological charge correlator 
in the continuum containing contact terms \cite{seiler:01}).\\

\section{Acknowledgements}

I would like to thank my collaborators over the years, namely 
P.\ van Baal,
E.\ Bilgici,
C.~Gattringer,
C.\ Hagen,
E.-M.\ Ilgenfritz, 
B.\ Martemyanov,
M. M\"uller-Preussker,
D.\ N{\'o}gr{\'a}di,
A.~Sch\"afer 
and S.\ Solbrig.
Support by DFG (BR 2872/4-1) is acknowledged.


\providecommand{\href}[2]{#2}\begingroup\raggedright\endgroup

\end{document}